# The Hall-Petch effect as a manifestation of the general size effect

Y. Li,[a] A.J. Bushby[b] and D.J. Dunstan[a*]

[a] School of Physics and Astronomy,
[b] School of Engineering and Materials Science,
Queen Mary University of London,
London E1 4NS, England.

The experimental evidence for the Hall-Petch dependence of strength on the inverse square-root of grain size is reviewed critically. Both the classic data and more recent results are considered. While the data can be fitted to the inverse square-root dependence excellently (but using two free fitting parameters for each dataset), it is also consistent with a dependence on the simple inverse of grain size (with one free fitting parameter for each dataset). There have been difficulties, recognised for half-a-century, in explaining the inverse square-root expression. A Bayesian analysis shows that the data strongly supports the simple inverse expression proposed. Since this expression derives from underlying theory, it is also more readily explicable. It is concluded that the Hall-Petch effect is not to be explained by the variety of theories found in the literature, but is a manifestation of, or underlain by, the general size effect observed throughout micromechanics, due to the inverse relationship between the stress required and the space available for dislocation sources to operate.

* Corresponding author, E-mail d.dunstan@qmul.ac.uk

## 1. Introduction

In the years around 1950, two effects of size were identified in the strength of materials; both can be summarised as *smaller is stronger*. Hall (1951) and Petch (1953) found that the strength of iron and steel increases when the grain size is smaller. On the basis of the theoretical work by Eshelby *et al.* (1951), their work established experimentally the eponymous relationship,

$$\sigma(d) = \sigma_0 + \frac{k_{HP}}{\sqrt{d}} \quad (1)$$

where $d$ is the grain size, $\sigma(d)$ is the stress at yield or a flow stress at higher plastic strains, $\sigma_0$ is the corresponding stress for large single crystals or very large-grained material (we refer to it here as the bulk stress), and $k_{HP}$ is a constant that may be predicted by theory or may be considered to be a material constant. This relationship was soon found to apply quite generally to other metals.

On the other hand, Frank and van der Merwe (1949) and later van der Merwe and co-workers, and especially Matthews and his co-workers, investigated theoretically and experimentally the elastic misfit strain that could be supported by thin epitaxial layers of one metal or semiconductor grown on another. By considering the force balance on threading dislocations or the minimum energy configuration of the system, Matthews developed the relationship between the maximum or critical thickness $h_c$ for a given misfit $\varepsilon_0$. In Matthews *et al.* (1970) this is given for an 001-oriented layer as

$$h_c = \frac{b(1 - \nu \cos^2 \theta)}{8\pi \varepsilon_0 (1 + \nu) \cos\lambda} \ln \frac{h_c}{b} \quad (2)$$

where $b$ is the magnitude of the Burgers vector, $\nu$ is the Poisson's ratio, and $\theta$ and $\lambda$ are angles between the slip plane and the Burgers vector and the growth plane. Many versions of Eq.2 were given subsequently by various authors (see Dunstan, 1997).

Over the decades that followed, these two size effects were addressed by different communities, with very little interaction. Matthews critical thickness theory was developed and applied within, largely, the semiconductor device community in the context of the strained heterostructures required for, e.g., semiconductor lasers (see Adams, 2011) and high-electron-mobility transistors (see O'Reilly, 1989). The theory remains essentially correct and the principal modification relevant here was the realisation that for significant plastic relaxation of the elastic strain a relaxation critical thickness needed to be defined, about four or five times the $h_c$ of Eq.2, to take account of the operation of dislocation sources (Beanland, 1992, 1995; Dunstan *et al.* 1996, Dunstan, 1997). From Eq.2, elastic strain rather than stress (i.e. stress normalised by the relevant elastic modulus), and normalised size (measured in units of Burgers vector or lattice constant), are the relevant parameters. These considerations lead to a general size-dependence equation,

$$\varepsilon_{el}(d) = \varepsilon_0 + \frac{k \ln d}{d} \quad (3)$$

where the dimensionless constant $k$ is expected to be of the order of unity, and in the context of the present paper $d$ will the normalised grain size. This equation is theoretically applicable to any situation where a dimension (such as grain size) constrains the size of the dislocation sources that have to operate if plasticity is to occur. We refer to it below as the size-effect equation from dislocation curvature (the DC equation).

Meanwhile, in the wider materials science community, a number of theories were put forward to supplement the Eshelby *et al.* (1951) theory in accounting for the Eq.1 description of the Hall-Petch effect and particularly the inverse-square root of *d* in it (see Section 4). Size effects became recognised in micro-mechanical testing generally, in nano-indentation (Nix and Gao, 1998), in thin wires under torsion (Fleck *et al*., 1994), in thin foils in flexure (Stölken and Evans, 1998), and most dramatically in micropillars under compression (Uchic *et al.* 2004). Despite a few key papers – such as that of Nix (1989) applying critical thickness theory to thin films, that of Thompson (1993) addressing grain size effects in thin films in the framework of critical thickness theory, and our own (Dunstan and Bushby, 2004) applying critical thickness theory to wire torsion and foil bending – theories of the micromechanical size effect proliferated in parallel with the various theories of the Hall-Petch effect. One symptom of this was the expression of the effect of the size of the specimen or of the loaded region in micromechanical testing as

$$\sigma(a) = \sigma_0 + ka^{-x} \qquad (4)$$

where *a* is some suitable characteristic dimension such as micropillar diameter or indentation contact radius. Much effort has been invested in finding appropriate values of the scaling exponent *x* for particular datasets, particular materials, for types of materials such as FCC or BCC metals, and for large collections of data (e.g. Korte and Clegg, 2010; Kraft *et al.*, 2010); however, we have suggested that such efforts are in vain. Despite apparent good fits to Eq.4 with various *x* in the range $0 < x < 1$, we reported that $x = 1$ (or $\ln d / d$) is better supported by the data (Dunstan and Bushby, 2013).

Returning to the Hall-Petch effect, many authors have considered exponents other than $x = \frac{1}{2}$. For example, Armstrong and Elban (2012) report that Mathewson (1919) fitted Eq.4 with $x = \frac{1}{4}$ to the data of Bassett and Davis (1919) for brass. Bragg (1942), before there was any significant body of experimental data, proposed $x = 1$, using the correct argument (repeated later by many authors in many contexts) that stresses are proportional to dislocation curvatures and therefore must scale inversely with size. Baldwin (1958), before $x = \frac{1}{2}$ was well-established, noted that experimentally it was hard to establish the true value of *x*, "If the experimentalist plots yield strengths against the corresponding inverses of the grain diameters and obtains a series of points falling on a straight line he becomes a partisan of Bragg's. If he obtains a straight line by plotting yield strengths against the corresponding inverses of the square roots of the grain diameters . . . *complete quote*. . ." the consequences are obvious. Kocks (1959) was aware of Baldwin's argument, and even after $x = \frac{1}{2}$ was well-established, Kocks (1970) still argued that the $x = 1$ and $x = \frac{1}{2}$ fits to data would be hard to distinguish. Unfortunately, his schematic illustration of this (his Fig.26) is rather unpersuasive. Nevertheless, he argued that "genuine grain size effects . . . are negligible except at very small grain sizes where they should be proportional to $1/d$ rather than $1/\sqrt{d}$". Hirth (1972) acknowledged these earlier proposals, but did not consider them further and adopted $x = \frac{1}{2}$ for the rest of his paper.

In the context of the relationship between subgrain or dislocation cell size and stress during work-hardening, Langford and Cohen (1970) argued for $x = 1$, using a model (expansion of a dislocation loop across a grain) that has the same physics as Bragg (1942) and the Matthews theory and the same formula, Eq.3, to within a numerical constant close to unity. Holt (1970) collected literature data for several pure metals and pointed out that the "results of individual investigator can, as many of them recognised, be fitted by a straight line of slope –1, i.e. $d \propto \sigma^{-1}$." Thomson (1970) concurred, and discussed the ambiguity of fitting datasets to find exponents – nevertheless stating explicitly that the Hall-Petch equation with $x = ½$ was well-established for high-angle grain boundaries. Kuhlmann-Wilsdorf and van der Merwe (1982) made the key formulation of the same physics, the argument of similitude, that if a dislocation structure is at equilibrium under a stress $\tau$, then if that structure is rescaled to a size $n$ times smaller, the stress must now be $n\tau$. This argument gives directly an exponent $x = 1$ (or $\ln d / d$) for the subgrain size-stress relationship in agreement with Langford and Cohen (1970). See also Kuhlmann-Wilsdorf (1970). Raj and Pharr (1986) collated a large amount of experimental data from the literature in which exponents ranging from 0 to 1 were reported. They identified a correlation between the prefactor and the exponent. This correlation was of the type that suggests that the range of fitting parameters is simply due to experimental error (see Yelon *et al*., 2012 and Dunstan, 1998). Raj and Pharr (1986) concluded that the best value from the data was $x = 0.84$, and in the light of the Kuhlmann-Wilsdorf and van der Merwe (1982) theory, the most plausible conclusion was indeed that $x = 1$. The theory is very closely related to the Matthews theory of critical thickness and if the logarithmic factor of Eq.2 is included, the theoretical value becomes ~0.9 (depending on the range of subgrain sizes considered). It may be thought surprising neither these nor subsequent papers (e.g. Kuhlmann-Wilsdorf and Hansen, 1991) considered extending the argument from sub-grains to the Hall-Petch grain-size effect too.

More recently, and particularly in the context of grain sizes below 10µm, there has been renewed interest in data and theories consistent with $x > ½$. Ohno and Okumura (2007) developed a theory based on the self-energy of geometrically necessary dislocations near a grain boundary from which can be obtained (combining their Eq.52 and Eq.56 and considering numerical values from their Fig.8) the expression of Eq.3 with $k$ of the order of unity. They also presented a collection of datasets from the literature in excellent agreement with this equation. Balint *et al*. (2008), using two-dimensional discrete dislocation dynamics (DDD) simulation of polycrystals, reported Hall-Petch behaviour with an exponent in good agreement with the experimental data collated by Ohno and Okumura (2007), with an exponent near $x = ½$, rising to greater than 1 for smaller crystals. Agraie-Khafri *et al*. (2012) fitted their data for steel with $x = 0.66$ (see Section 2.1.1). However, none of these authors questioned the applicability of $x = ½$ for larger grain sizes. Hansen (2004) concludes that although no mechanism has been quantified to the extent that it would verify Eq.1, nevertheless Eq.1 is empirical and has predictive capability. Saada (2005) remarked that if source operation controls the strength, the strength should be given by an expression of the form of Eq.3, but added, "which is not the case". Arzt (1998) also discussed the generality of Eq.3 in many manifestations of size effects, but again did not propose that the data considered to support Eq.1 in fact supports Eq.3.

In a previous paper we showed that the micromechanical data is consistent with Eq.4 with $x = 1$. With normalisation, it is also consistent with the expression of Eq.3 with $a$ for $d$ and with $k \sim 1$. We drew attention to the complete lack of any data falling *under* the line of Eq.3 in this form with *k* ~ 1 – we interpreted this by noting that other strengthening mechanisms may lead to data above the line, but if plasticity occurs through dislocation multiplication and motion, there are no weakening mechanisms to give data below the line. Eq.3 thus describes the *minimum* strength (Dunstan and Bushby, 2013).

It was then observed that a collection of datasets from the literature displaying the Hall-Petch effect are likewise concentrated above the line of the normalised Eq.3 with *k* ~ 1, and we proposed that this can be taken as experimental *support* for the applicability of Eq.3 to the Hall-Petch effect, while the data are merely *consistent* with Eq.1 (Dunstan and Bushby 2014). It is this proposal that we develop here. In Section 2 we review the data, both those used in Dunstan and Bushby (2014) and many additional datasets. In Section 3 we give a fully Bayesian analysis of the support the data gives to the different hypotheses, Eq.1 and Eq.3. Finally, in Section 4 we compare the predictions of the different theories of the Hall-Petch effect with the data. We conclude that Eq.3 and the theory from which it derives always apply. That is, it describes the size dependence of plasticity *in general*, underlying other effects which also increase the strength of metals.

## 2. Review and analysis of the data

### 2.1. *Data presentation and fits*

For each metal, we present the datasets as normally reported by the original authors (i.e. plots against the inverse square-root of grain size) changing only from the authors' units to SI units. For each dataset, our best fit to Eq.1 is shown, calculated using the *Mathematica*[©] function *NonlinearModelFit* with both $\sigma_0$ and $k_{HP}$ treated as free fitting parameters. It is not necessary to consider fits to Eq.2 with the scaling exponent *x* as a free fitting parameter – that issue was dealt with in Dunstan and Bushby (2014). Then we show the data plotted against the inverse of grain size, together with fits to Eq.3 with both $\sigma_0$ and *k* treated as free fitting parameters. Finally, we show the data normalised as for Eq.3 ($\sigma/Y$, $d/a_0$) on log-log plots, on which we show also the plot of Eq.3 with appropriate parameter values, $k = 0.72$ and $\sigma_0$ chosen so that the curve goes under the data – i.e. not fits, but interpretations of the expected minimum strength (Dunstan and Bushby, 2014).

For each dataset, we give what relevant information is available in the original papers about the metallurgical processing, especially grain-size modification and determination, and the yield or flow stress determination, or we mention the absence of this information. It should be noted that Rhines (1970) listed about ten ways of determining the grain size. For grains that are not equi-axed or which have a distribution of sizes, these methods can give diverse values for the average grain diameter. Since so few authors give their method of grain size determination, much of the variation in fitted values of *k* in Eq.3 or $k_{HP}$ in Eq.1 may arise from this.

### 2.1.1. *Iron and steel*

Hall (1951) and Petch (1953) both report data on iron and steel and it is appropriate therefore to begin with these metals. Sylwestrowicz and Hall (1951) measured wires under tension, made from three kinds of mild steel, Armco, Siemens-Martin (PXQ quality) and Basic Bessemer (Thomas quality) annealed *in vacuo* to obtain the different grain sizes. The method of measuring the grain sizes is not given. The lower yield points, measured at a strain rate of $10^{-4}\text{min}^{-1}$, were plotted by Hall (1951) and fitted by an inverse square-root dependence on grain size. We do not distinguish between the three kinds of mild steel in copying this data to Fig.1 (black crosses, see also the key in Fig.1c and the figure caption). While Hall did not plot the data against the inverse square-root of grain size, he followed Eshelby *et al.* (1951) in supposing that this should be the relationship between yield point and grain size and showed that his data (together with a datum for single-crystals from Holden and Hollonmon, 1949) were consistent with it.

Petch (1953) reported tensile experiment data on mild steel, ingot iron and spectrographic iron, indicated by black filled circles in Fig.1. The specimens ranged from single crystals to 8000 grains/sq. mm. They were prepared by annealing *in vacuo* at temperatures up to 1050°C for times up to 24h, followed by cooling at various rates. The single crystals and a few coarse-grained specimens were obtained by straining and annealing the decarburized mild steel. The method of measuring the grain sizes is described as "6–12 counts" on the surface, and to give not the true diameter but to be proportional to the true diameter. Petch was primarily concerned with the cleavage strength, which he showed obeyed Eq.1 and explained by identifying glide planes on which the dislocation motion is blocked by grain boundaries with Griffith cracks (Griffith, 1920) and again invoking the theory of Eshelby *et al.* (1951). However, he also plotted the lower yield point against $d^{-½}$ and fitted a single straight line to the data for all three metals, which accordingly we do not distinguish in Fig.1.

Armstrong *et al.* (1962) give data for the yield and flow stress of a 0.1%C semi-killed mild steel. The mild steel elemental composition is given. Grain sizes were obtained by 900–1200°C anneals, followed by machining and polishing and then a final anneal at 650°C. For the largest grains, annealing at 1200°C was followed by 8% strain and a long 650°C anneal. The method of measuring the grain sizes is not given. They find excellent agreement with linear dependences on $d^{-½}$, explained by invoking the Eshelby *et al.* (1951) theory, within the context of the Taylor (1938) theory of polycrystalline aggregates. In Fig.1 we reproduce their data (brown crosses) for the lower yield point, for 2.5% strain and 20% strain (intermediate strain values of 5%, 7.5%, 10% and 15% gave data evenly spaced between and parallel to the 2.5% and 20% data).

Douthwaite (1970) was interested in the relationship between hardness and flow stress, with the flow stress varied by different grain sizes and by strain-hardening. He studied EN2 steel (elemental compositions are given). Different grain sizes were obtained by annealing specimens in the range of 950°C-1250°C, followed by cooling at various rates through the critical range and a final anneal for 20h at 650°C. The largest grain size was obtained by strain annealing. The method of measuring the grain sizes is not given. All his data was plotted against $d^{-½}$ and given linear fits. In Fig.1 we plot his data for the flow stress of strained at 2.5%, 10% and 18.25% (green open circles).

Douthwaite and Evans (1973) studied 2.9% Si-steel with grain sizes varied by vacuum annealing between 650°C and 1150°C. Microstrain was obtained by tensile loading to below the yield point. Slip lines and grain boundaries were observed by optical microscopy after etching. The yield point was measured at 77K and 298K and they plotted their data with linear fits to $d^{-½}$. They followed Kelly (1966) in considering that elastic anisotropy can give rise to the $d^{-½}$ dependence. We plot their data in Fig.1 as red filled circles.

More recently, Kashyap and Tangri (1996) report tensile experiments at room temperature on 416L stainless steel. The specimens were vacuum annealed at temperatures from 900°C to 1300°C to obtain grain sizes from 3.1 to 104μm. Both optical and transmission electron microscopy were used to measure the grain size. Tensile testing was carried out with a strain rate of $1\times10^{-4}$ s$^{-1}$ at room temperature and at 400°C and 700°C. We plot their room temperature flow stress data (purple open circles) in Fig.1. Kashyap and Tangri were interested in the variation of the Hall-Petch parameters $\sigma_0$ and $k_{HP}$ of Eq.1 with strain. They considered the slip-distance model of Conrad (1963) and the anisotropy model of Ashby (1970), both of which predict $k_{HP} \propto \varepsilon^{½}$, but their data showed a much smaller strain sensitivity. Agraie-Khafri *et al*. (2012) also studied stainless, hot rolled AISI 301 sheet. After annealing for an hour at various temperatures 800–1200°C and forced-air-cooling, they measured grain sizes by both ultrasonic attenuation and by polishing and etching, and reported grain size distributions. Uniaxial tensile test data at 0.2% strain were plotted and fitted with $x = 0.66$. Their data is plotted in Fig.1 (red solid squares).

Some data for nanocrystalline iron (Embury and Fisher, 1966; Jang and Koch, 1990) are not considered here but are presented in Section 2.1.7 below.

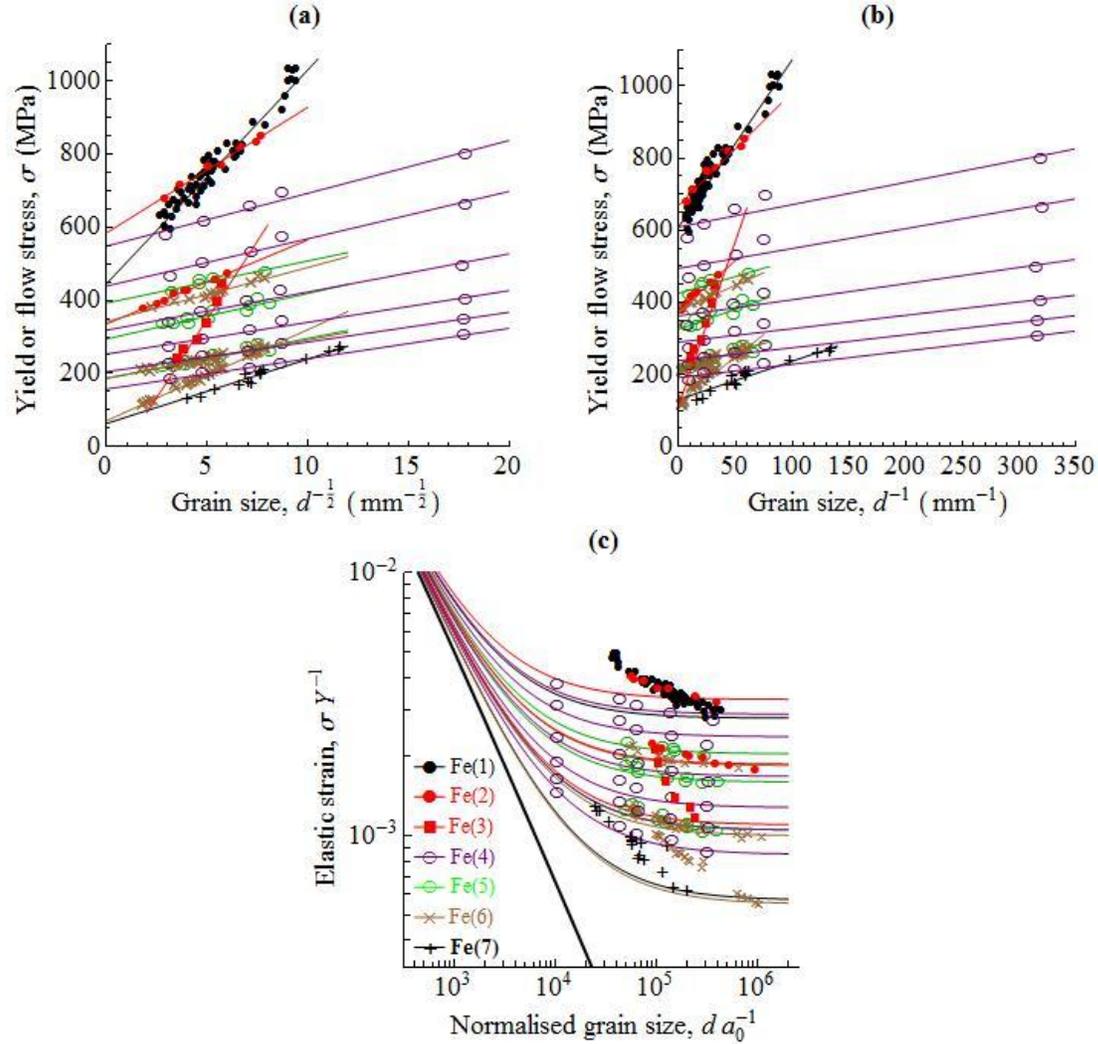

**Fig.1.** Datasets for iron, steel and silicon steel are plotted against (a) the inverse square-root of grain size, (b) the simple inverse of grain size, and (c) in normalised form on a double logarithmic plot. The classic datasets of Hall (1951) and Petch (1953) are indicated by black crosses, marked in the key in (c) as **Fe(7),** and filled circles, **Fe(1)**. The brown crosses, **Fe(6)**, represent data sets from Armstrong (1962). The green open circles, **Fe(5)**, come from the EN2 steel datasets in Douthwaite (1970). The red filled circles, **Fe(2)**, represent the silicon steel datasets from Douthwaite and Evans (1973) and the red solid squares, **Fe(3)**, are data from Agraie-Khafri *et al*. (2012). The Kashyap and Tangri (1996) data for 316L stainless steel are plotted as purple open circles, **Fe(4)**. In (a), the solid lines are fits to Eq.1; in (b) the solid lines are fits to Eq.4; in (c) the solid lines are plots of Eq.3 with $k = 0.72$ and $\sigma_0$ chosen so the lines for each dataset are below most of the data. The heavy black line is for Eq.3 with $\sigma_0 = 0$.

2.1.2. *Brass*

The oldest datasets come from Bassett and Davis (1919) and Babyak and Rhines (1960); neither of these papers presented plots of data against $d^{-\frac{1}{2}}$. Jindal and Armstrong (1967) replotted these datasets. Armstrong and Elban (2012) also reported that Mathewson (1919) fitted an inverse fourth-root to the data of Bassett and Davis. In Fig.2 we plot the Bassett and Davis data for 68-32 (red circles) and 69-31 (red crosses) brass and the Babyak and Rhines data for 70-30 brass (red triangles), taken from Jindal and Armstrong (1967).

Armstrong *et al*. (1962) also presented data for the flow stress of 70-30 brass at the yield point and at 2%, 5%, 10% and 20% strain. The grain size was varied by vacuum annealing at temperatures from 400°C to 800°C; how the grain size was measured was not stated. In Fig.2 we plot their data for the flow stress of 70-30 brass at yield stress and at 20% strain (blue crosses); the data at intermediate strains follows parallel regularly spaced lines.

Douthwaite (1970) also reports data for 70-30 brass. The grain size range was obtained by annealing for different times in the range 450-950°C. However, the method of measuring the grain sizes is not given. Tensile tests were carried out at a strain rate of $2\times10^{-4}$ s$^{-1}$. Data for the yield stress and the flow stress at 5% and 7.5% strain are given, and are plotted in Fig.2 (black open circles).

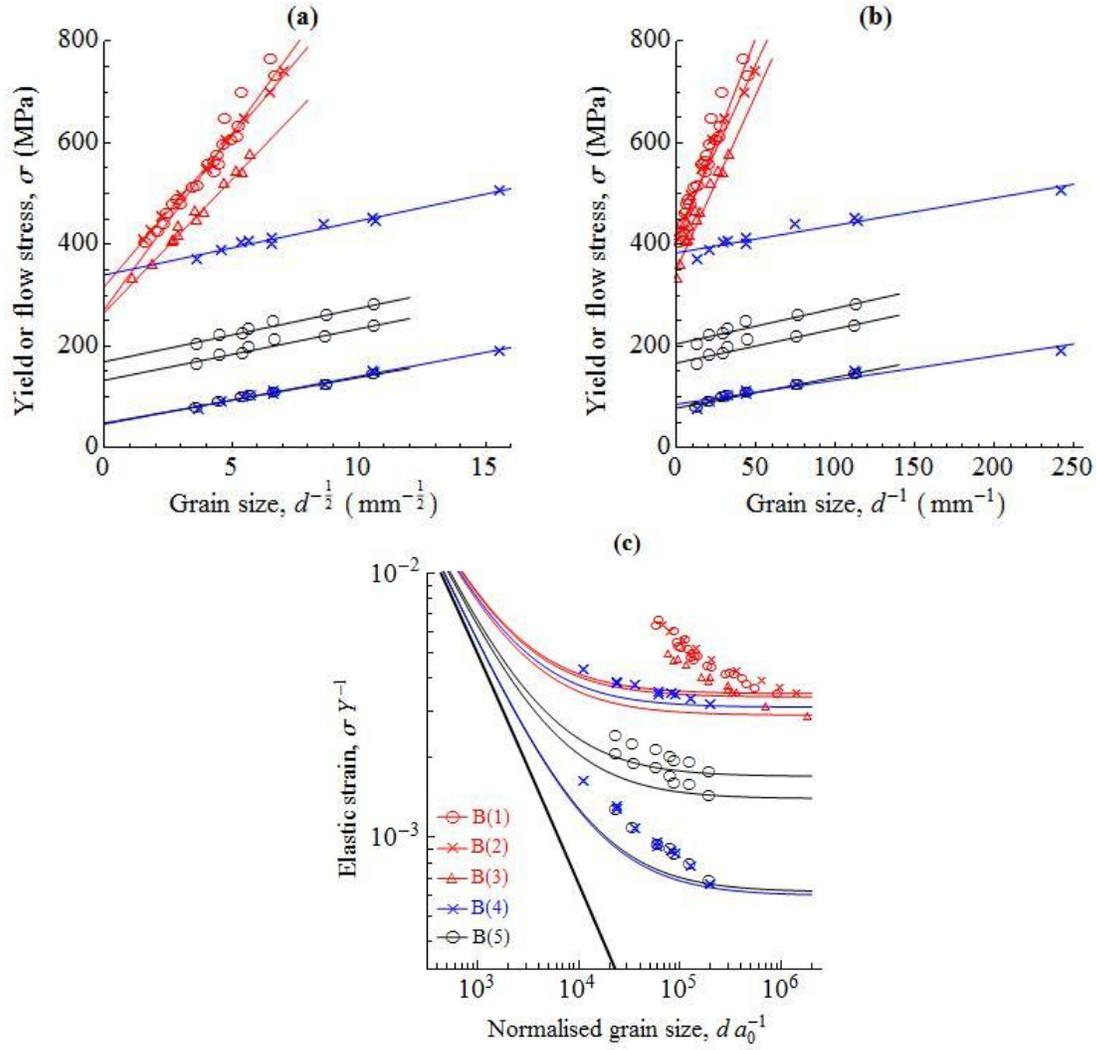

**Fig.2.** Datasets for brass are plotted against (a) the inverse square-root of grain size, (b) the simple inverse of grain size, and (c) in normalised form on a double logarithmic plot. The red circles, **B(1)**, crosses, **B(2)**, and triangles, **B(3)**, are the Bassett and Davis (1919) data for 68-32 and 69-31 brass and the Babyak (1960) data for 70-30 brass, respectively. The blue crosses, **B(4)**, are the Armstrong (1962) 70-30 brass data at yield stress and at 20% strain; the black circles, **B(5)**, are the data from Douthwaite (1970) for the yield stress and the flow stress at 5% and 7.5% strain. The solid curves are as in Fig.1.

### 2.1.3. *Copper*

Feltham and Meakin (1957) reported data for copper at 0.5% strain. Tensile specimens of oxygen-free copper (99.9911%) were annealed for various periods in vacuum, in the range 500–700°C. Grain size measurements are not described. Armstrong *et al*. (1962) plotted this data with a linear fit to $d^{-1/2}$, and we have replotted this data from Armstrong *et al*. in Fig.3 (green circles).

Hansen (1982) reported tensile test data for 99.999% copper. The copper was reduced by cold drawing and recrystallized at temperatures from 300-400°C for 1h. The recrystallized grain size was further increased by annealing at temperatures from 500-600°C. The grain size was determined by optical microscopy and specimens having grain sizes of 8.5, 25 and 60μm were examined. Tensile testing was carried out with a strain rate of $7\times10^{-4}\,\mathrm{s}^{-1}$, and after straining, detailed electron microscopy was carried out to elucidate the mechanisms of grain boundary hardening. We plot their data in Fig.3 for measurements at liquid nitrogen temperature and room temperature at 5%, 10% and 20% strain, indicated by black and red circles respectively.

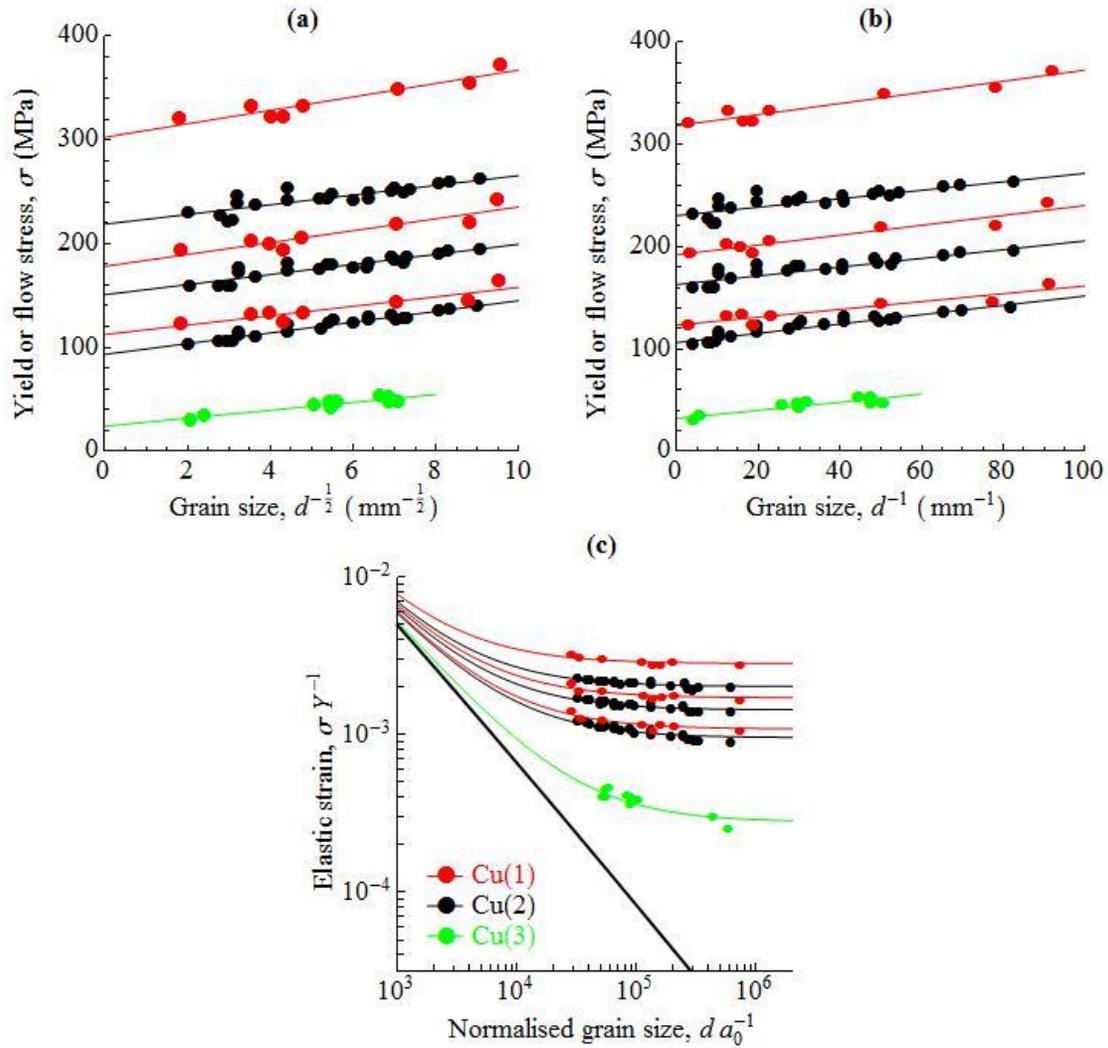

**Fig.3.** Data sets for copper are plotted against (a) the inverse square-root of grain size, (b) the simple inverse of grain size, and (c) in normalised form on a double logarithmic plot. The green circles indicate the data for copper at 0.5% strain from Armstrong et al. (1962), **Cu(3)**. Data from Hansen and Ralph (1982) are plotted, taken at room temperature, shown by black symbols, **Cu(2)** and liquid nitrogen temperature, shown by red symbols, **Cu(1)**, and at 5%, 10% and 20% strain, corresponding to the increased values of $\sigma_0$ required by the fits. The solid curves are as in Fig.1.

2.1.4. *Tungsten, titanium, chromium*

With few datasets per metal, we consider W, Ti and Cr together. Three papers give diamond point hardness (DPH) data; accordingly we divide by the Tabor factor 3 before normalising stress by *Y*. Vashi *et al*. (1970) consolidated 0.05μm tungsten powder material at a pressure of 1GPa for 10min at temperatures of 820$^o$C, 870$^o$C and 920$^o$C. The grain sizes ranged from 0.15μm to 10μm, measured by both optical metallography and electron microscopy. Hardness testing was performed with indentation sizes always large compared with the grain sizes. Their data is plotted in Fig.4 (black circles).

Chromium hardness data is given by Brittain *et al.* (1985). Different grain size specimens were prepared by electrodeposition. The method of measuring the grain sizes is not given. The diamond pyramid hardness values they report are plotted in Fig.4 (blue circles).

Hu and Cline (1968) report micro-hardness tests on titanium. The raw material was repeatedly cold-rolled and recrystallized by vacuum annealing. After the specimen preparation, the various grain sizes were obtained by a final anneal at various temperatures, and measured by both optical and transmission-electron microscopy. Their data is plotted in Fig.4 (red crosses).

Jones and Conrad (1969) reported tensile test data for alpha-Titanium at room temperature. Commercial purity (A-70) titanium was supplied as 1/4 inch diameter centreless-ground rod. These were cold-swaged to 0.078 inch diameter wire, and 2 inch lengths were recrystallized in a vacuum of $5\times10^{-5}$ torr to give grain sizes in the range 0.8 μm to 30μm. The grain size values were obtained from conventional optical techniques using the mean linear intercept method. Their data for the 4% strain flow stress are plotted in Fig.4 (red filled circles).

It is interesting to note that in Fig.4a and Fig.4b titanium appears to have a much weaker Hall-Petch effect than tungsten or chromium. However, this is just an effect of its much lower Youngs modulus, for in the normalised plot of Fig.4c it is not far away from tungsten or chromium.

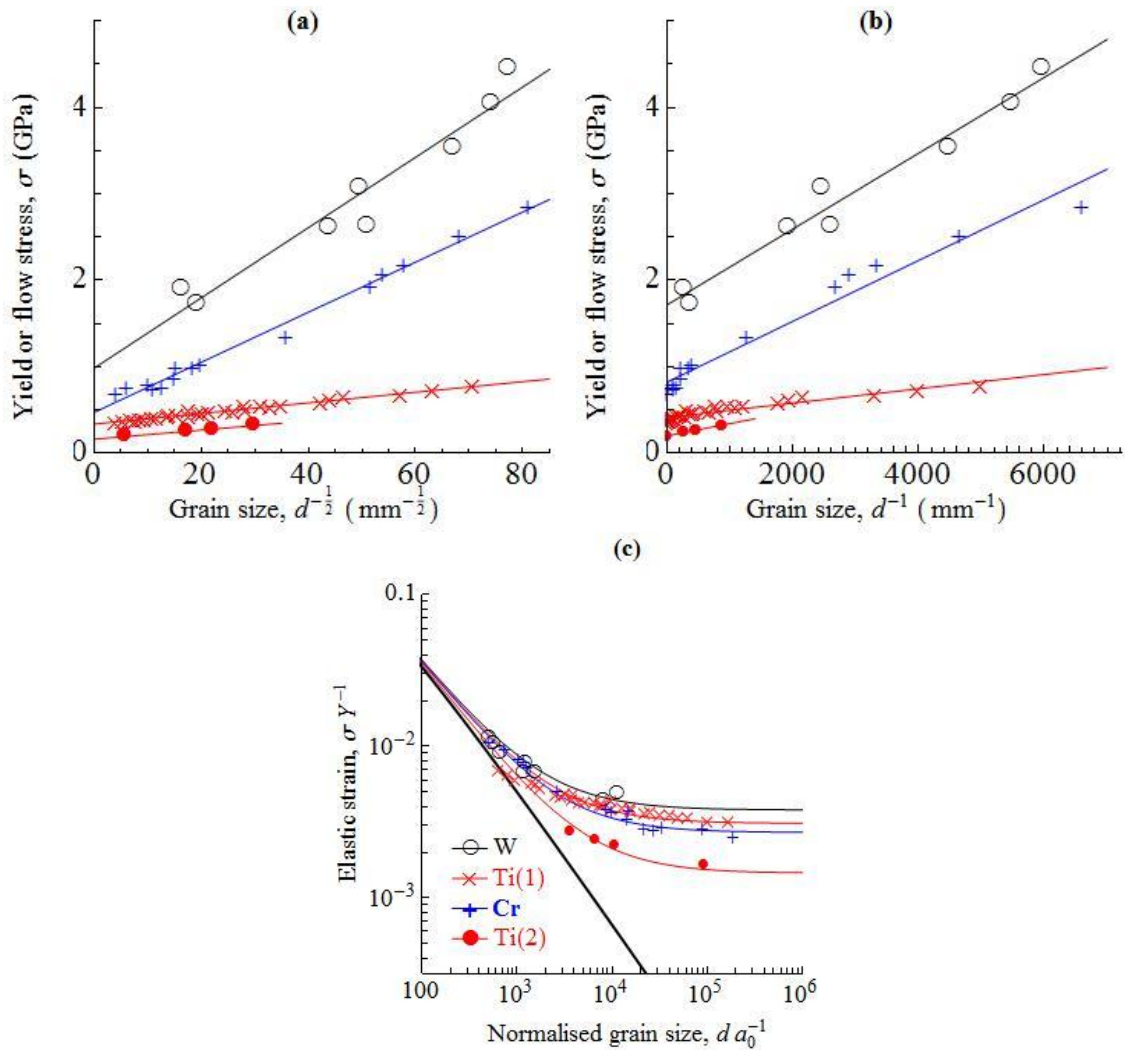

**Fig.4.** Datasets for tungsten, titanium and chromium are plotted against (a) the inverse square-root of grain size, (b) the simple inverse of grain size, and (c) in normalised form on a double logarithmic plot. The tungsten data, shown by black open circles, **W** in the key in (c), are taken from Vashi *et al.* (1970). Data for titanium are from Hu and Cline (1968) indicated by red crosses, **Ti(1)**, and from Jones and Conrad (1969) indicated by solid red circles, **Ti(2)**. The chromium data from Brittain *et al* (1985) are shown by blue crosses, **Cr**. The solid curves are as in Fig.1.

2.1.5. *Silver, Nickel*

Aldrich and Armstrong (1970) reported data for silver over a wide range of grain size. Specimens of cold-rolled sheet material of purity 99.9% were annealed in air at temperatures from 100°C to 900°C for ½hr. The grain size was measured from either optical and electron micrographs, and was taken as three-halves the mean linear intercept. Tensile testing was performed at 0.0667 min$^{-1}$ to fracture and they recorded the yield stress, the flow stress at 0.2%, 5% and 20% strain and the fracture stress. They compared linear fits to $d^{-1}$, $d^{-½}$ and $d^{-⅓}$ and concluded that $d^{-½}$ fitted best. Their data is shown in Fig.5 for yield stress (black circles) and flow stress at 20% strain (black crosses).

Thompson (1977) reported a study of work-hardening in nickel. Specimens with small grain size were obtained by electroplating from a high-purity bath and then annealing at various temperatures. Grain sizes of 80μm and larger came from swaged and annealed material, and a 6mm single-crystal specimen was also made. Grain size measurements are not described. Tensile tests were conducted at the strain rate of $8.3 \times 10^{-4}$ s$^{-1}$. No Hall-Petch plot is given but a log-log plot showed a slope of 0.367 for grain sizes above 1 μm. We plot Thompson's data for yield stress in Fig.5 (blue filled circles).

Keller and Hug (2008) reported tensile tests on nickel specimens with a thickness to grain size ratio *t*/*d* between 1.3 and 15. Specimens of 99.98% wt.% purity were annealed in vacuum for 220 min at temperatures between 600°C and 1050°C and then air cooled. The grain size and weak texture were revealed by electron back-scattered diffraction (EBSD). Uniaxial tensile tests were carried out at a strain rate of $2.4 \times 10^{-4}$ s$^{-1}$. At yield stress, they observed a normal Hall-Petch behaviour for all *t*/*d* values and this data is reproduced in Fig.5 (blue open circles). For higher strain and smaller *t*/*d*, deviations from the normal Hall-Petch behaviour were observed, and explained in terms of the effect of free surface on the work-hardening mechanisms. This data is not considered here.

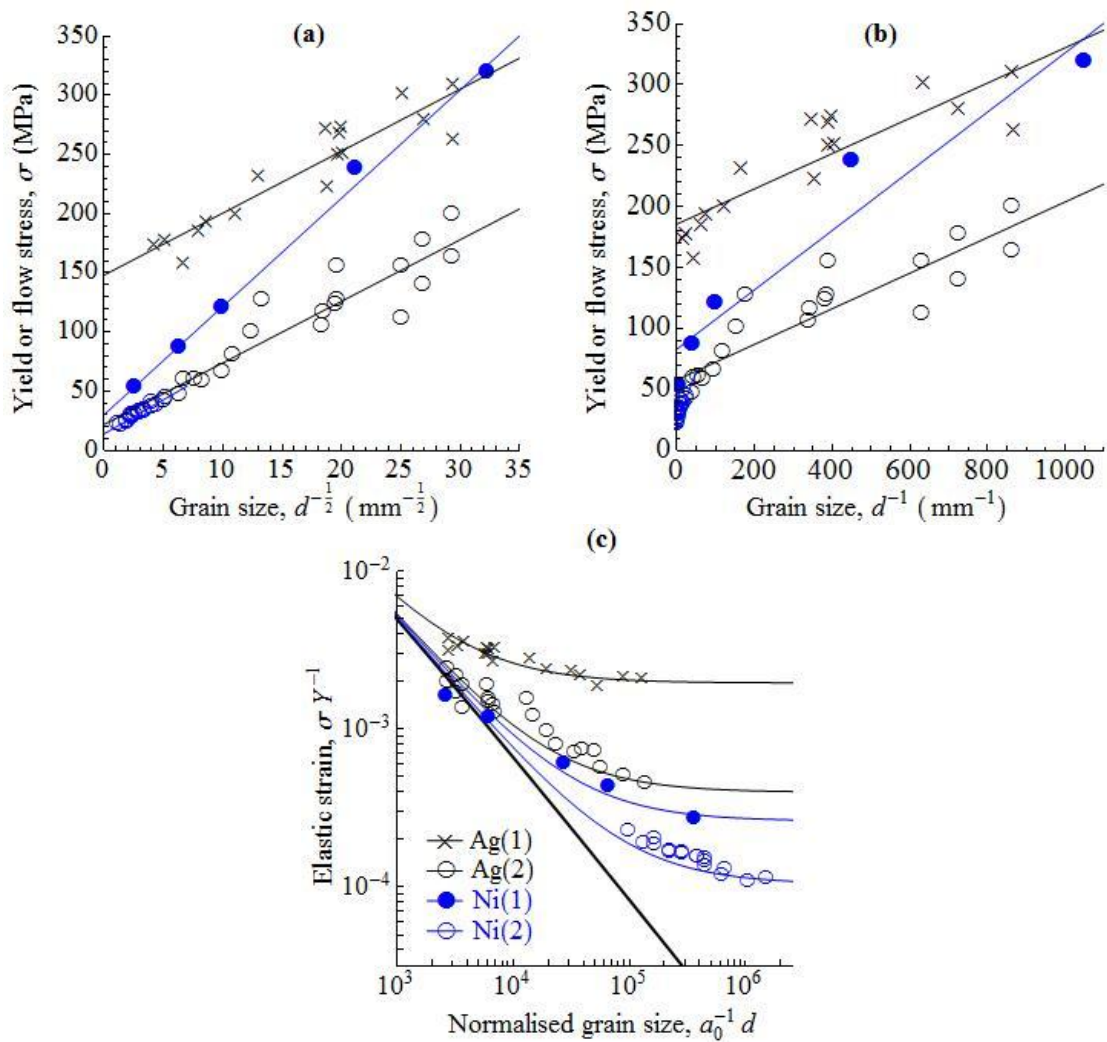

**Fig.5.** Data for silver and nickel are plotted against (a) the inverse square-root of grain size, (b) the simple inverse of grain size, and (c) in normalised form on a double logarithmic plot. The black circles and crosses indicated the silver data of Aldrich and Armstrong (1970) at yield stress, **Ag(2)**, and 20% strain, **Ag(1)**. Blue circles show the nickel data; the open circles, **Ni(2)**, the Keller and Hug (2008) data and the filled circles, **Ni(1)**, the Thompson (1977) data. The solid curves are as in Fig.1.

### 2.1.6. *Aluminium*

Carreker and Hibbard (1955) reported tensile tests on wires made from two batches of aluminium both of 99.987% purity (elemental compositions given). Samples were annealed in air for 1h at temperatures from 300°C to 500°C. Grain sizes were measured by optical microscopy. Tensile tests were carried out at a strain rate of $0.66 \times 10^{-4}\,\text{s}^{-1}$ at various temperatures down to 20K. They presented log-log plots; at 300K, 5% strain the flow stress scarcely depends on grain size while at 1% strain the slope is close to $-½$. We reproduce their data for 300K at 1% strain in Fig.6 (red crosses); their data for 0.5% strain is very similar.

Hansen (1977) reported tensile test data for two grades of aluminium, purities 99.999% and 99.5%. The material was reduced by cold rolling or drawing and annealed at temperatures from 300°C to 625°C. The recrystallized grain sizes were determined by optical microscopy in polarized light. Tensile tests were carried out at a strain rate of $0.66 \times 10^{-4}\,\text{s}^{-1}$. After the tests, the grain sizes of the specimens at 0.5%, 5% and 10% strain were measured by electron microscopy. The data for five strains from 0.2% to 20% are reproduced in Fig.6 (blue and black open circles for the low-purity and high-purity materials respectively).

Tsuji *et al.* (2002) prepared ultrafine-grained (UFG) aluminium by accumulative roll-bonding (ARB) process. Commercial purity aluminium (JIS-1100) was used for the ARB process. The material was annealed for 600 s or 1800 s at temperatures from 373 K to 673 K. The resulting grain sizes were measured by TEM, using the mean interception method. Tensile tests were carried out at a strain rate of $8.3 \times 10^{-4}\,\text{s}^{-1}$, and a linear fit to $d^{-½}$ was found for the yield point. This data is reproduced in Fig.6c (green triangles).

Yu *et al.* (2005) prepared ultrafine grained aluminium by equal channel angular extrusion of commercial purity aluminium, followed by annealing. For materials of larger grain sizes, the grain sizes were measured by EBSD. For materials of fine grain size, the grains were determined from TEM measurements. Tensile tests were conducted at a strain rate of $7.1 \times 10^{-4}\,\text{s}^{-1}$, at room temperature and 77K. They reported a good fit to the inverse-square root dependence (Eq.1) for grain sizes greater than 1μm. These data are reproduced in Fig.6c (room temperature, purple squares; 77K, circles).

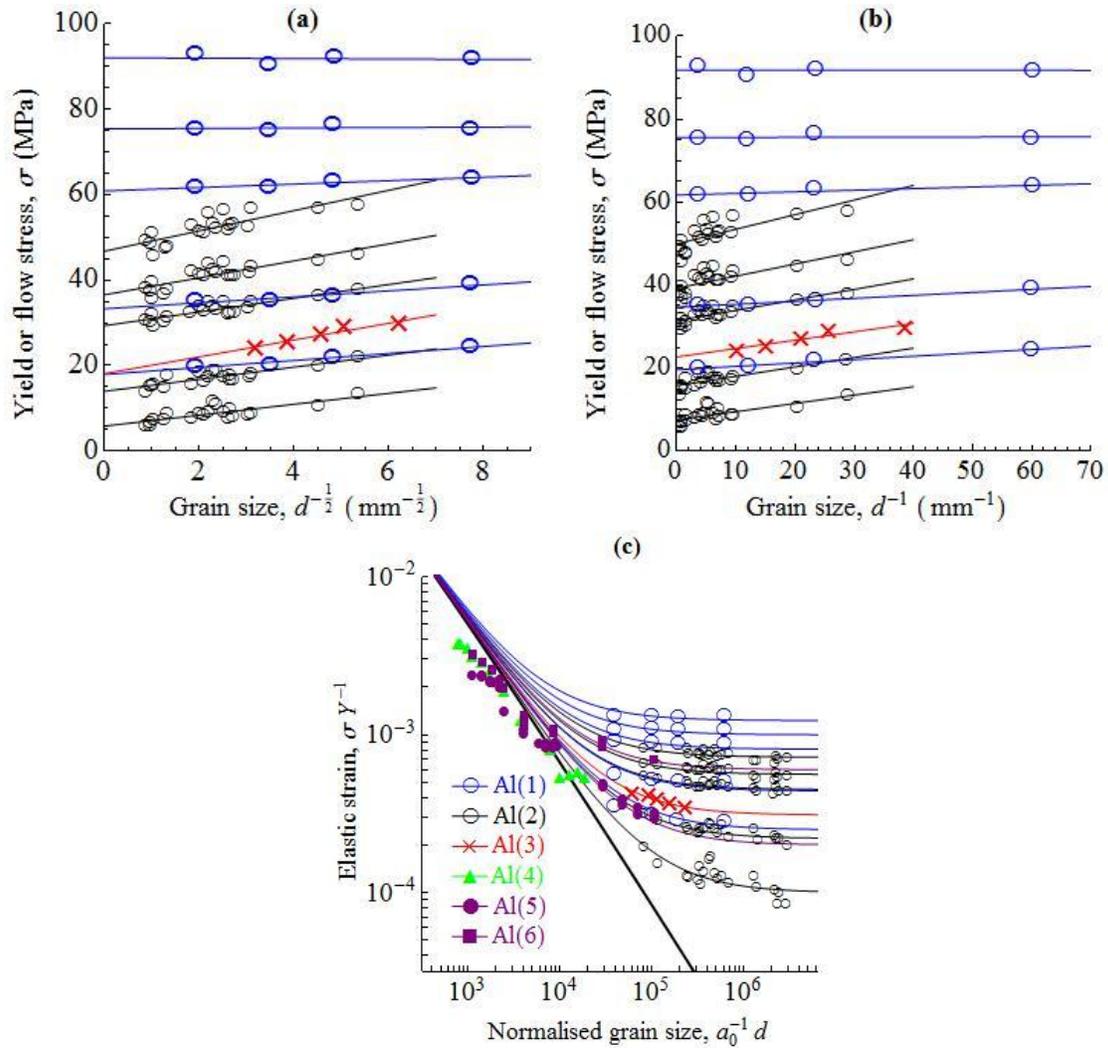

**Fig.6.** Data for aluminium are plotted against (a) the inverse square-root of grain size, (b) the simple inverse of grain size, and (c) in normalised form on a double logarithmic plot. The red crosses, **Al(3)**, show the Carreker and Hibbard (1957) aluminium data. The black and blue circles are from Hansen (1977), **Al(2)** for 99.999% and **Al(2)** for 99.5% aluminium at 0.2, 1, 5, 10, 20% strain at room temperature. The data for ultrafine grained material are shown only in (c) because of the very large values of $1/d$. The green solid triangles, **Al(4),** are the data of Tsuji *et al.* (2002) and the purple filled squares, **Al(6),** and circles, **Al(5),** come from Yu *et al.* (2005) at room temperature and at 77K respectively. The solid curves are as in Fig.1.

2.1.7. *Nanocrystalline metals*

As the grain size is reduced and the strength due to dislocation source operation increases towards the theoretical strength of the material, other mechanisms of plasticity may come into play and give a strength below that of Eq.3. Mechanisms such as grain-boundary sliding and diffusion have been considered in connection with the inverse Hall-Petch effect (e.g., Schiøtz *et* al., 1999; Wolf *et al*., 2003). A strength below that of Eq.3 will generally be reached at a larger grain size than a strength below that of Eq.1. Accordingly, although the inverse Hall-Petch effect is outside our scope here and while we do not review comprehensively the literature on the strength of nanocrystalline metals, it is interesting to compare a few datasets in this range with Eq.3.

Data for nano-crystalline iron has been reported by Embury and Fisher (1966) and by Yang and Koch (1990). These data are plotted in Fig.7 in the region where they fall significantly below fits using Eq.3. For comparison, the Ti data of Hu and Cline (1968) from Fig.3 and the Al data of Tsuji *et al.* (2002) from Fig.6 are plotted in the same region. We see that all four datasets begin to fall below their Eq,3 fits at about the same normalised grain size of $a_0^{-1}d \sim 3000$.

2.2. *Discussion*

Comparison of the fits of Fig.1-7 (a) with the fits of Fig.1-7 (b) by eye suggests that these fits to the data are almost equally good. That is, unless a rigorous statistical analysis were to show otherwise (and no such analysis has been given in the literature) there is almost no experimental evidence that Eq.1 is correct; that the strength of metals varies as the inverse square-root of grain size. This comparison supports Baldwin's (1958) comment; if Hall, and then Petch, and subsequent authors, had used a $d^{-1}$ abscissa and had a theoretical argument to hand supporting a $d^{-1}$ dependence, then most likely that is what would now be known as the Hall-Petch equation. That is not to say that a $d^{-1}$ dependence is supported by Figs 1-7 (b) any more than an inverse square-root dependence is supported by Figs 1-7 (a). The data are consistent with both, but the data support neither dependence. Indeed, from these plots, there is not even any support for a power-law as in Eq.3. The best that can be said is that the strength increases monotonically as the grain size is decreased.

The plots of Fig.1-7 (c) offer a different insight. The heavy line of Eq.3 with $\sigma_0 = 0$ clearly divides the plot into two regions. Almost all the data is found above the line. This is the same result as we found for micromechanical testing data (Dunstan and Bushby, 2013), and we propose the same interpretation. This line indicates a minimum strength, as a function of size, which is consistent with the theory of dislocation source operation. Then data can be found on this line. Data can also be found above it and this is expected, when there are other strengthening mechanisms in play (e.g. Peierls stress, dislocation starvation, Taylor or forest hardening, etc). Some of these mechanisms may also be correlated with grain size, and that possibility permits us to account for the data (particularly for iron and brass) that slope above our Eq.3 fits in the log-log plots. Only at small *d* are any data found appreciably below this line, and this is expected when weakening mechanisms that are not dislocation-based come in to play, such as grain boundary sliding.

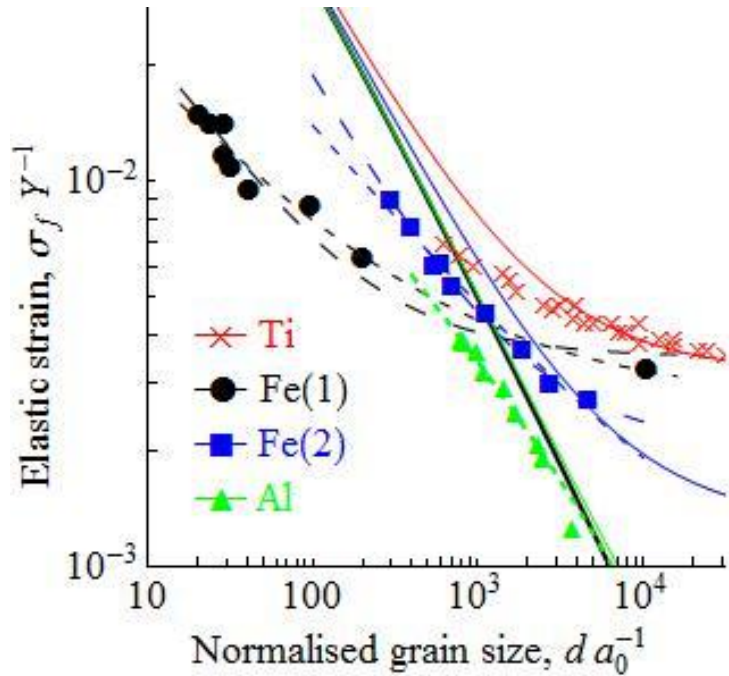

**Fig.7.** Data for some nanocrystalline metals. Nanocrystalline iron is shown by blue open squares, **Fe(2),** from Embury and Fisher (1966) and black filled circles, **Fe(1)**, from Jang and Koch (1990). For aluminium, the data of Tsuji *et al.* (2002) is copied from Fig.6 (green solid triangles, **Al**) and for titanium the data of Hu and Cline (1968) is copied from Fig.4 (red crosses, **Ti**). The heavy black line is from Eq.3 with $\sigma_0 = 0$. The solid curves for each data set are from Eq.3 with suitable values of $\sigma_0$. The short-dashed lines for the nc-Fe data are the fits using Eq.1 and the long-dashed curves, Eq.4.

    All the datasets presented agree excellently with Eq.1, but it is noteworthy that, following the original authors, the fits to Eq.1 treat both $\sigma_0$ and $k_{HP}$ as free fitting parameters. This is reasonable for $\sigma_0$, for the bulk strengths of metals can generally be varied very widely by metallurgical processing. It is less reasonable for $k_{HP}$ – it might have been expected that this would have a fixed value given by theory. Failing that, it might have been hoped to be a material constant. Instead, it varies widely even for each individual metal where there is more than one dataset. It is treated as a free fitting parameter.

    In contrast, many of the datasets presented agree quite well with Eq.3 for the fixed value of $k = 0.72$. In cases where the grain size and material treatment are known more fully, e.g. Figs 3 and 4, the interpretation given by Eq.3 is excellent.

## 3. Bayesian analysis of support for hypotheses

Previously, we reported an analysis of the statistical support that the data provides for the different theories. We used the semi-Bayesian approach of calculating the likelihood $L$ of the data under the different theories – which was inconclusive – and then taking into account the Akaike information criterion (AIC), which provides a heavy weighting against theories with more free fitting parameters. Since the Hall-Petch theory of Eq.1 has two free fitting parameters per data set ($\sigma_0$ and $k_{HP}$; 34 parameters for 17 datasets) while the theory underlying Eq.3 has only one free fitting parameter per data set ($\sigma_0$) plus one ($k \sim 1$) for all datasets (18 parameters for 17 datasets), the AIC gave odds of many millions to one that the dislocation curvature theory is true and the Hall-Petch theory false (Dunstan and Bushby, 2014).

Here we give a more fundamental, fully Bayesian analysis. Bayes' theorem may expressed in the form,

$$\frac{P(H|\text{new data})}{P(\text{not } H|\text{new data})} = \frac{P(H|\text{old data})}{P(\text{not } H|\text{old data})} \times \frac{P(\text{new data}|H)}{P(\text{new data}|\text{not } H)} \quad (5)$$

that is, the new odds on the hypothesis under test ($H$) being true when new data is acquired are the old odds from before, times the ratio of the probability of the new data under the hypothesis $H$ and its probability if $H$ is false. Here, we may take $H$ to be the hypothesis that Eq.3 is valid, and its negation to be the hypothesis that Eq.1 is valid

Eq. 5 may be applied very directly to our problem. In the absence of a theory constraining the values of $\sigma_0$ and $k_{HP}$ in Eq.1, the experimentally-determined values of yield or flow stress against grain size are expected to have a uniform probability distribution in the $\log\sigma_0 - \log d$ space, by Benford's Law (Newcomb, 1881; Benford, 1938). Let this probability distribution be represented by the relative value 1 everywhere. On the other hand, Eq.3 with $\sigma_0 = 0$ divides the $\log\sigma_0 - \log d$ space into two equal parts – one below the $1/d$ line (with or without the $\ln d$ term makes no significant difference) and one above. Eq.3 asserts that the probability that data will be below the line is close to zero, so the data should be concentrated into the half of the space above the line. So, defining $H$ as the hypothesis that Eq.3 is correct, we have a relative probability density of about 2 for data above the $1/d$ line and nearly 0 for data below the $1/d$ line.

We apply Eq.5 iteratively for each dataset. We start with a value for the first term on the RHS. This term expresses the prior probability $P_0$ or odds $O_0$ that $H$ is true. Using only the Principle of Insufficient Reason, we would take $P_0 = \frac{1}{2}$, making this term 1 (even odds) On the other hand, we might consider that the probability that an equation that has stood for sixty years is false is low, so that perhaps we should take $P_0 = 10^{-3}$. The first dataset, that falls above the $1/d$ line, gives a value 2 to the second term on the RHS (odds of 2:1), so that the term on the LHS, the odds on $H$ become 2:1. This becomes the first term of the RHS ($O_1 = 2$, $P_1 = \frac{2}{3}$) when we consider the second dataset. If each successive dataset $i$ falls above the $1/d$ line, we see at once that $O_i = 2 O_{i-1}$ and for $n$ datasets $O_n = 2^n P_0$. So just ten or twenty such datasets give overwhelming odds on $H$, depending on whatever reasonable prior $P_0$ we may have chosen. Here we have 55 datasets, giving odds of $2^{55} P_0$ – which is overwhelming for any reasonable choice of prior.

These odds on the Eq.3 hypothesis can be reduced slightly by considering that not all the datasets are independent. If the data for the yield point of a metal fall on and above the line, it is predictable that the data for the same material after work-hardening will also fall above, so the observation that this is so does not strengthen the hypothesis. However, where the yield-point data fall on an Eq.3 ($\sigma_0 > 0$) line, it is not predictable that the work-hardened data would also fall on an Eq.3 line with just a higher value of $\sigma_0$, as observed e.g. in Fig.6. These considerations reduce the number of independent datasets to about 45, which still leaves overwhelming odds on *H*.

## 4. One-parameter Hall-Petch theories

While experimentally Eq.1 is treated as if both σ$_0$ and $k_{HP}$ are free fitting parameters, the theories which have been put forward to account for the inverse square-root law of Eq.1 do of course make predictions for $k_{HP}$. And the phenomena in question, when they occur in practice, must contribute to the strength. It is appropriate, therefore, to compare their predictions with the data, to test whether they are in fact supported by the data. The four classic theories of the Hall-Petch inverse-square-root dependence on *d* (Eq.1) are shown schematically in Fig.8(a-d) together with schematic representations of the dislocation curvature theory leading to Eq.3 in epitaxial layers (Fig.8e) and in polycrystalline metals (Fig.8f).

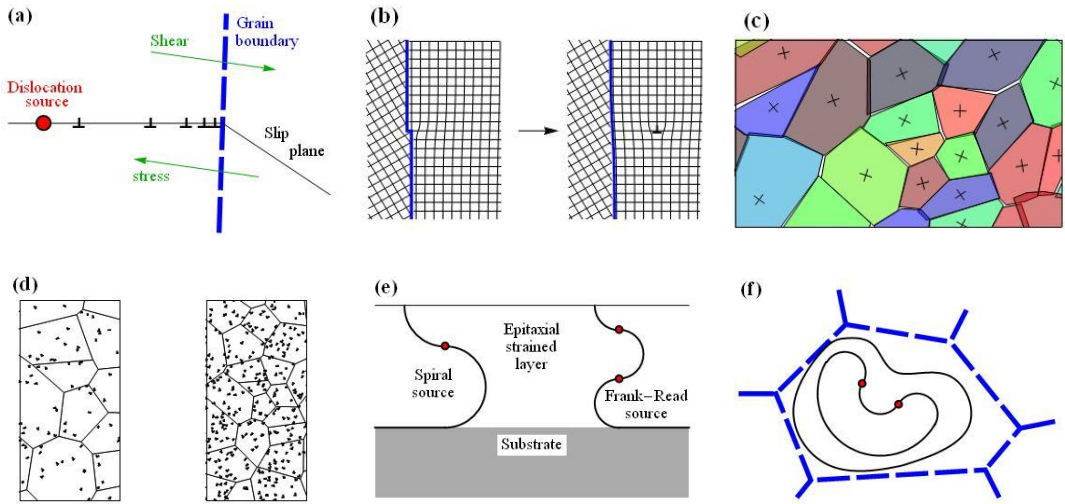

**Fig.8.** The four classic models are shown schematically (a–d) together with the dislocation curvature model (e, f). In (a), the pile-up theory of Eshelby *et al.* (1951) is represented. The ledge-emission model of Li (1963) is shown in (b). The effect of crystalline anisotropy in forcing stress and strain gradients (Kelly, 1966; Ashby, 1970) is illustrated in (c), in which anisotropic grains are subject to a homogenous stress field. In (d), the dislocation densities resulting from the slip-distance theory (Conrad, 1963; Kuhlmann-Wilsdorf, 1970) are shown as low in large grain and high in small grain material. The Matthews critical thickness concept is illustrated in (e) for the form well-established in semiconductor epitaxy, for the spiral source and the Frank-Read source in a strained epitaxial layer on a substrate, and in (f) a Frank-Read source is shown in a single grain to illustrate that smaller grains require greater curvatures and hence stresses varying as 1/*d* or ln*d* / *d*.

Simulation and modelling are beginning to be able to display the Hall-Petch effect and predict Hall-Petch slopes (e.g. Lim *et al.*, 2011, 2014). However, to the extent they match experimental results, they are currently no more able than the experimental results to determine the correct value(s) of the exponent *x*. For that reason we do not consider these methods further.

4.1. *Dislocation pile-up model*

This is the phenomenon perhaps most often proposed to account for the Hall-Petch Eq.1. In this model, a dislocation source in a grain operates many times under an applied stress to produce a number of dislocations on the same glide plane (Fig.8a). The leading dislocation experiences a force from the stress field, and also the forces from the following dislocations behind it, but it is blocked from further movement by the grain boundary. When the force on the leading dislocation is sufficient to stress the material at or beyond the grain boundary beyond theoretical strength (or some lower value), dislocations are produced in the neighbouring grain and large-scale plasticity becomes possible. Following Cottrell (1949), Eshelby *et al.* (1951) and Antolovich and Armstrong (2011), the theory gives

$$\tau = \tau_0 + \sqrt{\frac{Gb\tau_c}{2\pi d}} \qquad (6)$$

where $\tau_c$ is the critical shear stress at the grain boundary at which a dislocation is generated in the neighbouring grain. The maximum reasonable value of $\tau_c$ is the theoretical strength, less than $G/10$. Taking $\tau/G \sim \sigma/Y$ and $b \sim a_0$, Eq.6 becomes

$$\varepsilon_f - \varepsilon_0 \leq \frac{\sim 0.1}{\sqrt{a_0^{-1} d}} \qquad (7)$$

for the elastic flow strain $\varepsilon_f = \sigma_f Y^{-1}$, in normalised units as used in Fig.1c-7c and Fig.9.

The data from Section 2 are compared with the prediction of the pile-up model in Fig.9. The shaded triangle below the solid line is the allowed region according to Eq.7. Many of the datasets have slopes (values of $k_{HP}$) greater, even very much greater, than the predictions. Applying the same statistics as in Section 3, the odds *against* the pile-up model appear to be greater than the odds *on* the hypothesis $H$ that Eq.3 is correct; for many data are falling where their relative probability is much less than one-half. In fact this is an exaggeration. Different data sets falling where their probabilities are low are evidently not independent events. A Bayesian analysis of this problem must start with the *a priori* estimate of the (small) probability $P_0$ that the model is correct but the parameter values in Eq.6 have been wrongly estimated. Then all data wherever they fall are fully consistent with this hypothesis and the hypothesis retains the probability $P_0$ independent of any data.

Pile-up can of course occur, and will give rise to some (grain-size dependent) strengthening. However, Fig.9 shows that it cannot account for the most part of the strength in most datasets; that is, it is a weak effect compared with the direct effect of grain size on the dislocation mechanisms that are required for plasticity (source operation, Eq.3). This conclusion is confirmed by discrete dislocation dynamics simulations of wires in torsion, in which pile-up can be encouraged by prohibiting cross-slip or reduced by allowing cross-slip. The simulations showed torque-torsion curves that did not change significantly with the amount of pile-up (Senger *et al.*, 2012).

### 4.2. *Grain boundary ledge model*

Li (1963) sought a model that could explain the Hall-Petch behaviour in the majority of cases where there is no evidence of dislocation pile-up. He proposed that grain boundaries and sub-grain boundaries should emit dislocations (Fig.8b). Then he showed that the stresses required are nearly the same (1) for a pile-up to drive a dislocation through a grain boundary, (2) for a pile-up to activate a source on the other side of the boundary, and (3) to move dislocations in a forest formed by all the dislocations emitted by a tilt boundary. In model (3) the grain-size dependence arises from the density of the forest. Murr (1975) reported observations by electron microscopy supporting this model. The prediction of this model is

$$\sigma - \sigma_0 = \alpha G b \sqrt{\frac{8m}{\pi d}} \tag{8}$$

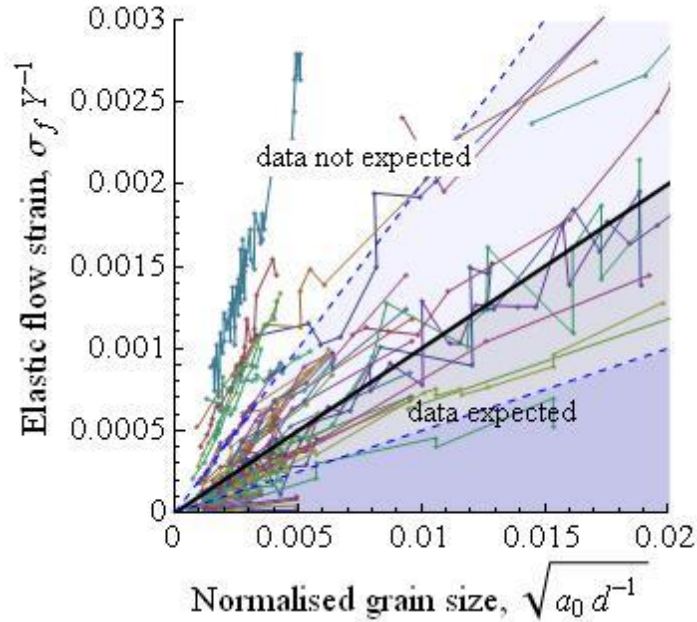

**Fig.9.** The predictions of the pile-up model (Eq.7) (heavy black line) and the grain-boundary ledge model (Eq.9) (dashed black lines indicating the range of the upper limit of the grain-boundary edge model predictions) are compared with the data of Section 2. The depth of shading indicates schematically the probability according to these models that data will fall in the various regions; white corresponds to a probability close to zero.

where $\alpha$ is a constant of the order of 0.4, and the constant $m$ is given by the expectation that the product $mb$ will be in the range 0.02 to 0.2 (Chia *et al.*, 2005; Conrad and Jung, 2005; Zhu *et al.*, 2008). Converting to $\varepsilon = \sigma/Y$ and using $b \sim a_0$, this becomes,

$$\varepsilon - \varepsilon_0 \sim \frac{0.05 - 0.15}{\sqrt{a_0^{-1} d}} \tag{9}$$

Fig.8 shows the range where data is to be expected according to Eq.9. This model is again inconsistent with much of the data, and it does not account for the wide scatter of the data. The same considerations apply to the probability that it is correct as they do to the pile-up model in Section 4.2.

4.3. *Slip-distance model*

Conrad (1963) and co-workers (1967, 2005) developed a theory which gives very naturally the inverse square-root dependence on grain size of Eq.1 when square-root work-hardening occurs. See also Kuhlmann-Wilsdorf (1970). Mobile dislocations account for the plastic strain $\varepsilon_{pl}$, and

$$\varepsilon_{pl} = \rho_m b \bar{x} \tag{10}$$

where $\rho_m$ is the density of mobile dislocation and $\bar{x}$ is their mean free path, taken as proportional to the grain size, $\bar{x} = \lambda d$ It is further assumed that a constant proportion of the total dislocation density $\rho$ is mobile, so that $\rho_m = \xi\rho$. Using the Taylor (forest) hardening expression, substituting and rearranging, we have

$$\sigma - \sigma_0 = \alpha G b \sqrt{\rho} = \alpha G \sqrt{\frac{b \varepsilon_{pl}}{\lambda \xi d}} \tag{11}$$

where $\alpha$ is the Taylor coefficient. Normalising as previously and taking $G = \frac{1}{2}Y$

$$\varepsilon - \varepsilon_0 = \frac{1}{2}\alpha \sqrt{\frac{\varepsilon_{pl}}{\lambda \xi}} \sqrt{a_0 d^{-1}} \tag{12}$$

Thus the model attributes the Hall-Petch effect to the increased dislocation density in small grains (Fig.8c) due to the reduced slip distance. This gives a Hall-Petch coefficient which vanishes at the yield point ($\varepsilon_{pl} = 0$) and is proportional to the square-root of plastic strain otherwise. The constant $\alpha$ is normally taken as about 0.3, while the constants $\lambda$ and $\xi$ are both of the order of but less than unity, so the factor $\alpha^2 / \lambda\xi$ may be taken to be about unity. Then for the datasets reported at high plastic strains ~0.2, the value of $k_{HP}$ in Eq.12 may be close to unity, while for the datasets reported near the yield point ($\varepsilon_{pl}$ ~ 0.002) it will be closer to 0.1. These two possibilities are plotted on Fig.8 (chain-dotted lines). Clearly, this theory can be ruled out for the Hall-Petch effect near the yield point, but it survives as a candidate for explaining Eq.1 behaviour at high plastic strains when inverse-square-root work-hardening is observed.

4.4. *Elastic anisotropy model*

This model goes back to Kelly (1966) and Ashby (1970). Given the random orientation of grains, a homogenous stress field necessitates an inhomogeneous strain field, resulting in gaps and overlaps between grains as shown in Fig.8d. Here a two-dimensional polycrystalline cubic material with $c_{11} = 1$, $c_{12} = \frac{1}{2}$, $c_{44} = \frac{1}{2}$, normalised anisotropy $C = |c_{11} - c_{12} - 2c_{44}|/Y = \frac{2}{3}$ is shown under the uniform stress field $((0, \sigma), (\sigma, 0))$ with $\sigma = 0.1$. Deforming the grains to eliminate the gaps and overlaps results in inhomogeneous stress and strain fields – and the resulting strain gradients require geometrically necessary dislocations and a consequent increase in strength. The grain size dependence arises naturally, in that if the grains are smaller the strain gradients and the densities of GNDs will be proportionately larger. This model predicts that under suitable normalisation $k_{HP}$ will be proportional to the elastic anisotropy. In Fig. 10(a) we plot the values of $k_{HP}$ for the cubic metals against the normalised anisotropy parameter $C/Y$. While there is a considerable scatter of the data for metals where we have more than one value, it is clear that there is no strict dependence, nor even a trend suggesting that $k_{HP}$ depends upon $C$. This model is therefore neither consistent with nor explanatory of the data.

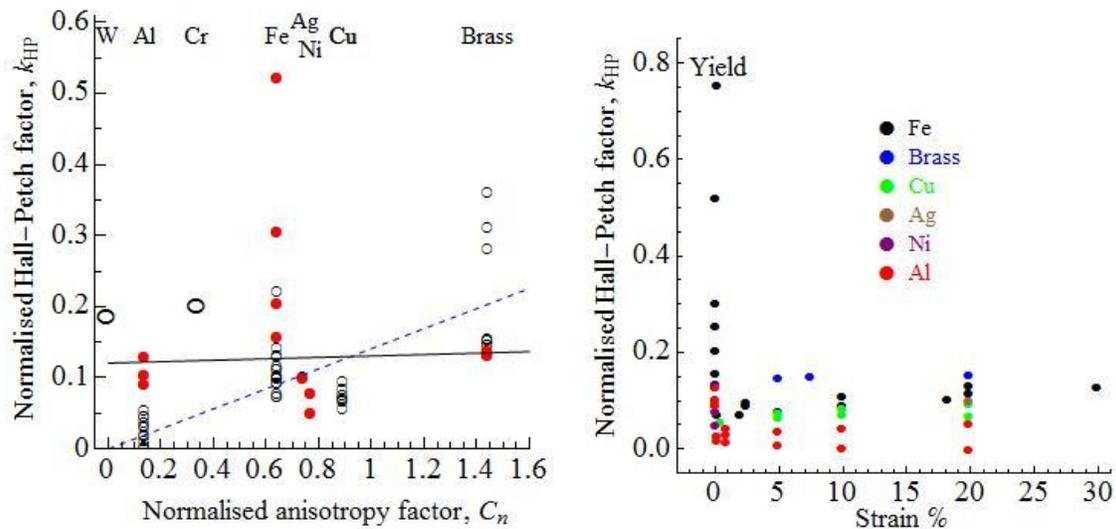

**Fig. 10.** In (a), the Hall-Petch coefficients are plotted against the normalised anisotropy factor as described in the text. The large data points indicate that only one data set is available for a metal; the small data points represent many results for the same metal. The red filled circles indicate the yield stress data sets. The latter are averaged to give a single value for each metal before fitting. The solid black line is a least-squares fit of $y = ax + b$ to the averaged data and the dashed blue line a fit of $y = ax$ as described in Section 4.4. The dependence of the Hall-Petch coefficient on the plastic strain is shown in (b). For all the data sets where data is reported for different flow strains, the data shows no evidence of any dependence on strain.

# 5. Conclusions

It is clear that there is neither experimental nor theoretical evidence for the 60-year-old Hall-Petch equation, Eq.1. In particular the wide range of experimentally reported values for the Hall-Petch constant, $k_{HP}$, for similar materials do not support Eq. 1, neither are they predicted by any of the theories in section 4. On the other hand, the large body of data is consistent with the size-effect expected from dislocation curvature, Eq.3, for the minimum strength expected for a given grain size. That consistency depends on the necessary caveats: non-dislocation based plasticity such as grain-boundary sliding may take over at small grain sizes; other strengthening mechanisms may be correlated with grain size without being caused by grain size.

An argument in favour of this conclusion is that it brings the Hall-Petch effect under the umbrella of the size effect(s) generally, rather than being *sui generis* with its own unique inverse-square-root exponent and therefore a need for its own explanations. We argue that the underlying size-dependence that dictates the minimum strength for dislocation plasticity should be in the singular – the only size effect that will be necessarily present in all experimental situations is the Orowan mechanism, applied to this context by Bragg (1942): size must be inversely proportional to dislocation curvature and hence to stress.

It might be considered that this doesn't matter. It might be pointed out that the Hall-Petch relation, Eq.1, is a valid empirical relation and as such it is useful for prediction – for interpolation and extrapolation of material properties – whether or not it is theoretically correct. That is certainly so for interpolation, for which it will be as useful – but no more useful – than a smooth curve drawn through the data by hand – but it is a very dangerous approach to extrapolation.

One may also regret the loss of time – the wasted effort – in attempts to explain the inverse-square-root form of Eq.1, and of course the parameter values therein too. On the other hand, one may anticipate theoretical and practical advances that may be made when it is considered that the grain-size effect operates through the same mechanism as other size effects and therefore may be combined with them (as in Ehrler coupling of structural size and grain size; see Dunstan *et al.*, 2009).

The other main conclusion from this work is that it can never be sufficiently strongly emphasised that a good fit of data to an equation or a theory is of no significance unless it has been adequately considered what else might fit the data. This is a much more general conclusion, of interest to non-metallurgists as much as to metallurgists.

**Acknowledgements**
Too many colleagues have contributed valuable insights and comments to name them all. But one must be singled out.  We are very grateful to Prof. Ron Armstrong, who appears frequently in the list of references, for his assistance in accessing the older literature, and especially for his encouragement and his enthusiasm for a re-appraisal of a topic with which he has been intimately concerned for over fifty years.

YL is grateful to the Chinese Scholarship Council for his PhD studentship.

**References**

Adams, A.R., 2011. Strained-layer quantum-well lasers. IEEE J. Sel. Topics Quantum Electronics 17, 1364–1373.

Aghaie-Khafri, M., Honarvar, F., Zanganeh, S., 2012. Characterisation of grain size and yield strength in AISI 301 stainless steel using ultrasonic attenuation measurements. J. Nondestruct. Eval. 31, 191–196.

Aldrich, J.W., Armstrong, R.W., 1970. The grain size dependence of the yield, flow and fracture stress of commercial purity silver. Metall. Trans. 1, 2547 -

Antolovich, S.D., Armstrong, R.W., 2014. Plastic strain localization in metals: Origins and consequences. Prog. Mater. Sci. 59, 1–160.

Armstrong, R., Codd, I., Douthwaite, R.M., Petch, N.J., 1962. The plastic deformation of polycrystalline aggregates. Phil. Mag 7,  45–58.

Armstrong, R.W., Elban, W.L., 2012. Hardness properties across multiscales of applied loads and material structures. Mater. Sci. Technol. 28, 1060–1071.

Arzt, E., 1998. Size effects in materials due to microstructural and dimensional constraints: A comparative review. Acta Mat. 46, 5611–5626.

Ashby, M.F., 1970. The deformation of plastically non-homogeneous materials. Phil. Mag. 21, 399–424.

Babyak, W.J., Rhines, F.N., 1960. The relationship between the boundary area and hardness of recrystallized cartridge brass. Trans. TMS-AIME 218, 21–23.

Baldwin, W.M., 1958. Yield strength of metals as a function of grain size. Acta Metall. 6, 139–141.

Balint, D.S., Deshpande, V.S., Needleman, A., Van der Giessen, E., 2008. Discrete dislocation plasticity analysis of the grain size dependence of the flow strength of polycrystals. Int. J. Plasticity 24, 2149–2172.

Bassett, W.H., Davis, C.H., 1919. Comparison of grain-size measurements and Brinell hardness of cartridge brass. Trans. TMS-AIME 60, 428–449.

Beanland, R., 1992. Multiplication of misfit dislocations in epitaxial layers. J. Appl Phys. 72, 4031–4035.

Beanland, R, 1995. Dislocation multiplication mechanisms in low-misfit strained epitaxial layers. J. Appl Phys. 77, 6217–6222.

Benford, F., 1938. The law of anomalous numbers. Proc. Am. Philos. Soc. 78, 551–572.

Bragg, L., 1942. A theory of the strength of metals. Nature 149, 511–513.

Brittain, C.P., Armstrong, R.W., Smith, G.C., 1985. Hall-Petch dependence for ultrafine grain size electrodeposited chromium. Scripta Metall. 19, 89–91.

Carreker, R.P., Hibbard, W.R., 1955, Wright Air Development Center Technical Report 55-113.

Chia, K.-H, Jung, K., Conrad, H., 2005. Dislocation density model for the effect of grain size on the flow stress of a Ti–15.2 at% Mo β-alloy at 4.2–650 K. Mater. Sci. Eng. A409, 32–38.
Conrad, H., 1963. Effect of grain size on the lower yield and flow stress on iron and steel. Acta Met. 11, 75–77.
Conrad, H., Feuerstein, S., Rice, L., 1967. Effects of grain size on the dislocation density and flow stress of niobium. Mater. Sci. Eng. 2, 157–168.
Conrad, H., Jung, K., 2005. Effect of grain size from millimetres to nanometers on the flow stress and deformation kinetics of Ag. Mater. Sci. Eng. A391, 272–284.
Cottrell, A.H., Bilby B.A., 1949. Dislocation theory of yielding and strain ageing of iron. Proc. Phys. Soc. London A62, 49–62.
Dollar, M., Thompson, A.W., 1987. The effect of grain size and strain on the tensile flow stress of quenched aluminium. Acta Met. 35, 227–235.
Douthwaite, R.M., 1970. Relationship between the hardness, flow stress, and grain size of metals. J. Iron Steel Institute 208, 265–269.
Douthwaite, R.M. and Evans, J.T., 1973. Microstrain in polycrystalline aggregates. Acta Met. 21, 525–530.
Dunstan, D.J., 1997. Strain and strain relaxation in semiconductors. J. Mater. Sci.: Mater. in Electronics **8**, 337–375.
Dunstan, D.J., 1998. The role of experimental error in Arrhenius plots: Self-diffusion in semiconductors. Solid State Commun. 107, 159–163.
Dunstan, D.J., Bushby, A.J., 2004. Theory of deformation in small volumes of materials. Proc. Roy. Soc. A460, 2781–2796.
Dunstan, D.J., Bushby, A.J., 2013. The scaling exponent in the size effect of small scale plastic deformation. Int. J. Plasticity 40, 152–162.
Dunstan, D.J., Bushby, A.J., 2014. Grain size dependence of the strength of metals: The Hall–Petch effect does not scale as the inverse square root of grain size. Int. J. Plasticity 53, 56–65.
Dunstan, D.J., Kidd, P., Beanland, R., Sacedón, A., Calleja, A., González, L., González, Y., Pacheco, F.J. 1996. Predictability of plastic relaxation in metamorphic epitaxy. Mater. Sci. Technol. 12, 181–186.
Dunstan, D.J., Ehrler, B., Bossis, R., Joly, S., P'ng, K.M.Y., Bushby, A.J., 2009. Elastic limit and strain hardening of thin wires in torsion. Phys. Rev. Lett. 103, 155501.
Embury, J.D., Fisher, R.M., 1966. The structure and properties of drawn pearlite. Acta Metall. 14, 147–159.
Eshelby, J.D., Frank, F.C., Nabarro, F.R.N., 1951. The equilibrium of linear arrays of dislocations. Phil Mag. 42, 351–364.
Feltham, P., Meakin, J.D., 1957. On the mechanism of work hardening in face-centred cubic metals, with special reference to polycrystalline copper. Phil. Mag. 2, 105–112.
Fleck, N.A., Muller, G.M., Ashby, M.F., Hutchinson, J.W., 1994. Strain gradient plasticity: Theory and experiment. Acta Metal. Mater. 42, 475–487.
Frank, F.C., van der Merwe, J.H., 1949. One-dimensional dislocations. II. Misfitting monolayers and oriented overgrowth. Proc. Roy. Soc. A198, 216–225.
Griffith, A.A., 1920. The phenomena of rupture and flow in solids. Phil. Trans. Roy. Soc. A 221, 163–198.

Hall, E.O., 1951. The deformation and ageing of mild steel: III Discussion of results. Proc. Phys. Soc. B 64, 747–753.

Hansen, N., 1977. The effect of grain size and strain on the tensile flow stress of aluminium at room temperature. Acta Metall. 25, 863–869.

Hansen, N., 2004. Hall-Petch relation and boundary strengthening. Scripta Mat. 51, 801–806.

Hansen, N., Ralph, B., 1982. The strain and grain size dependence of the flow stress of copper. Acta Met. 30, 411–417.

Hirth, J.P. 1972. The influence of grain boundaries on mechanical properties. Met. Trans. 3, 3047–3067.

Holt, D.L., 1970. Dislocation cell formation in metals. J. Appl. Phys. 41, 3197–3201.

Hu, H., Cline, R.S., 1968. Mechanism of reorientation during recrystallization of polycrystalline titanium. Trans. TMS-AIME. 242, 1013–1024.

Jang, J.S.C., Koch, C.C., 1990. The Hall–Petch relationship in nanocrystalline iron produced by ball milling. Scripta Metall. Mater. 24, 1599–1604.

Jindal, P.C., Armstrong, R.W., 1967. The dependence of the hardness of cartridge brass on grain size. Trans. TMS-AIME 239, 1856–1857.

Jones, R.L. and Conrad, H., 1969. Trans. TMS-AIME 245, 779. Data given in Abbaschian, R.and Reed-Hill, R.E., 2010. Physical Metallurgy Principles (Cengage Learning), p.192.

Kashyap, B.P., Tangri, K., 1997. Hall-Petch relationship and substructural evolution in boron containing type 316L stainless steel. Acta Mat. 45, 2383–2395.

Keller, C., Hug, E., 2008. Hall-Petch behaviour of Ni polycrystals with a few grains per thickness. Mater. Lett. 62, 1718–1720.

Kelly, A., 1966. Strong Solids (Clarendon Press, Oxford) p.84.

Kocks, U.F., 1959. Comments on "Yield strength of metals as a function of grain size". Acta Metall. 7, 131.

Kocks, U.F., 1970. The relationship between polycrystal deformation and single-crystal deformation. Met. Trans. 1, 1121–1143.

Korte, S., Clegg, W.J., 2010. Discussion of the dependence of the effect of size on the yield stress in hard materials studied by microcompression of MgO. Phil. Mag. 91, 1150–1162.

Kraft, O., Gruber, P.A., Mönig, R., Weygand, D., 2010. Plasticity in confined dimensions. Annual Review of Material Research 40, 293–317.

Kuhlmann-Wilsdorf, D., 1970. A critical test on theories of work-hardening for the case of drawn iron wire. Metall. Trans. 1, 3173–3179.

Kuhlmann-Wilsdorf, D., van der Merwe, J.H., 1982. Theory of dislocation cell sizes in deformed metals. Mater. Sci. Eng. 55, 79–83.

Kuhlmann-Wilsdorf, D., Hansen, N., 1991. Geometrically necessary, incidental and subgrain boundaries. Scripta Met. Mat. 25, 1557–1562.

Langford, G., Cohen, M., 1970. Calculation of cell-size strengthening of wire-drawn iron. Met. Trans. 1, 1478–1480.

Li, J.M.C., 1963. Petch relation and grain boundary sources. Trans. TMS 227, 239–247.

Lim, H., Subedi, S., Fullwood, D.T., Adams, B.L., Wagoner, R.H., 2014. A practical mesoscale polycrystal model to predict dislocation densities and the Hall-Petch effect. Mater. Trans. 55, 35–38.

Lim, H., Lee, M.G., Kim, J.H., Adams, B.L., Wagoner, R.H., 2011. Simulation of polycrystal deformation with grain and grain boundary effects. Int. J. Plasticity 27, 1328–1354.

Matthews, J.W., Mader, S., Light, T.B., 1970. Accommodation of misfit across the interface between crystals of semiconducting compounds or elements. J. Appl. Phys. 41, 3800–3804.

Mathewson, 1919. Discussion. Trans. TMS-AIME 60, 451–455.

Murr, L.E., 1975. Some observations of grain boundary ledges and ledges as dislocation sources in metals and alloys. Met. Trans. 6A, 505–513.

Newcomb, S., 1881. Note on the frequency of use of the different digits in natural numbers. Am. J. Math. 4, 39–40.

Nix, W.D., 1989. Mechanical properties of thin films. Metall. Trans. 20A, 2217–2245.

Nix, W.D., Gao, H., 1998. Indentation size effects in crystalline materials: A law for strain gradient plasticity. J. Mech. Phys. Sol. 46, 411–425.

Ohmo, N., Okumura, D., 2007. Higher-order stress and grain size effects due to self-energy of geometrically necessary dislocations. J. Mech. Phys. Solids 55, 1879–1898.

O'Reilly, E.P., 1989. Valence band engineering in strained-layer structures. Semicon. Sci. Technol. 4, 121–137.

Petch, N.J., 1953. The cleavage strength of polycrystals. J. Iron Steel Institute 174, 25–28.

Raj, S.V., Pharr, G.M., 1986. A compilation and analysis for the stress dependence of the subgrain size. Mater. Sci. Eng. 81, 217–237.

Rhines, F.N., 1970. Geometry of grain boundaries. Met. Trans. 1, 1105–1120.

Saada, G., 2005. Hall-Petch revisited. Mat. Sci. Eng. A400-401, 146-149.

Schiøtz, J., Vegge, T., Tolla, F.D.D., Jacobesen, K.W., 1999. Atomic-scale simulations of the mechanical deformation of nanocrystalline metals. Phys. Rev. B 60, 11971-11983.

Senger, J., Weygand, D., Dunstan D.J., 2012. Private communication (unpublished); work presented at Plasticity 2014 and TMS 2013.

Stölken J.S., Evans, A.G., 1998. A microbend test method for measuring the plasticity length scale. Acta Met. 46, 5109–5115.

Sylwestrowicz, W., Hall, E.O., 1951. The deformation and ageing of mild steel. Proc. Phys. Soc. B 64, 495–502.

Taylor, G.I., 1938. *Polycrystalline aggregates*. J. Inst. Met. 62, 307–???

Thompson, C.V., 1993. The yield stress of polycrystalline thin films. J. Mater. Res. 8, 237–238.

Thompson, A.W., 1970. Substructure strengthening mechanisms. Metall. Trans. 8A 833–842.

Thompson, A.W., 1977. Effect of grain size on work hardening in nickel. Metallurgica 25, 83–86.

Tsuji, N., Ito, Y., Saito, Y., Minamino, Y., 2002. Strength and ductility of ultrafine grained aluminum and iron produced by ARB and annealing. Scripta Mater. 47, 893–899.

Uchic, M.D., Dimiduk, D.M., Florando, J.N., Nix, W.D., 2004. Sample dimensions influence strength abd crystal plasticity. Science 305, 986–989.

Vashi, U.K., Armstrong, R.W. Zima, G.E., 1970. The hardness and grain size of consolidated tungsten powder. Metall. Trans. 1, 1769–1771.

Wolf, D., Yamakov, V., Phillpot, S.R., Mukherjee, A.K., 2003. Deformation mechanism and inverse Hall–Petch behavior in nanocrystalline materials. Zeitschrift für Metallkunde 94, 1091–1097.

Yelon, A., Sacher E., Linert, W., 2012. Comment on ''The mathematical origins of the kinetic compensation effect'' Parts 1 and 2 by P. J. Barrie, Phys. Chem. Chem. Phys., 2012, 14, 318 and 327. Phys. Chem. Chem. Phys. 14, 8232–8234.

Yu, C.Y., Kao, P.W., Chang, C.P., 2005. Transition of tensile deformation behaviors in ultrafine-grained aluminium. Acta Mat. 53, 4019–4028.

Zhu, T.T., Bushby, A.J., Dunstan, D.J., 2008. Materials mechanical size effects: a review. Mater. Technol. 23, 193–209.